# Mesoscale investigations of fluid-solid interaction: Liquid slip flow in a parallel-plate microchannel

Zi Li,[a, b, *] Jiawei Li,[b] Guanxi Yan,[b] Sergio Galindo-Torres,[a] Alexander Scheuermann,[b] and Ling Li[a]

[a] School of Engineering, Westlake University, Hangzhou, China
[b] School of Civil Engineering, The University of Queensland, Brisbane, Australia
[*] Corresponding author. Email address: lizi@westlake.edu.cn (Zi Li)

## Abstract

Liquid slip flow with a Knudsen number Kn = 0.001–0.1 plays a dominant role in confined flow channels. Its physical origin can be attributed to the intermolecular fluid-solid (F-S) interaction force. To this end, we propose herein continuous force functions (decaying either exponentially or by a power law) between fluid particles and two confined flat walls in the framework of the mesoscopic lattice Boltzmann model (LBM). The analytical solutions for density profile, velocity profile, slip length, and permeability ratio are derived and related to the mesoscale F-S interaction parameters and the size of the gap of the flow channel. Through nondimensionalization of the analytical solutions, we obtain the dimensionless numbers that indicate the key feature of slip-flow systems for each of the proposed force functions. The analytical solutions are strictly consistent with the LBM solutions. We suggest reasonable ranges for the F-S interaction parameters based on the observed range of density ratio (film fluid to bulk fluid) and the increasing permeability ratios with narrowing gap size. Within the given range of interaction parameters, simple relationships between permeability ratios and dimensionless numbers are provided by fitting. The curves for continuous F-S interaction force with two free parameters are calibrated for a hydrophobic surface by using LBM simulations, which were validated *a priori* by comparison with the slip velocity profile measured in a benchmark flow experiment. The mesoscopic LBM model based on the proposed F-S interaction force functions provides a robust framework to elucidate the physical process of liquid slip flow.

## 1. Introduction

The phenomenon of fluid slip has long been known in continuum fluid dynamics. Because the slip length quantifying the slip (defined as the distance between the fictitious solid wall where the bulk velocity extrapolates to zero and the real solid wall) is usually as little as 1–100 nm [1], the no-slip condition works well for macroscale flows. Slip flow, however, plays a dominant role in microfluidic devices. This paper focuses on the liquid flow in the slip regime for a Knudsen number Kn = 0.001–0.1 [2] (Kn is defined as the ratio of the molecular mean-free path to the characteristic length scale). Measurements have revealed enhanced flow rates in thin capillaries [3, 4] and the velocity profile in hydrophobic microchannel flows [5, 6]. Even the adsorbed water layer [7] and lower-than-expected water flow [8] can be observed in low-permeability reservoirs. Therefore, the fluid slip (damping) should be accounted for with care when estimating the transport properties of a liquid in tight porous media [9].

Ruckenstein and Rajora [10] proposed that fluid slip occurs over a gap (rather than directly on solid walls) that is generated by a release of entrained or soluble gas. Derjaguin and Churaev [11] attributed the slip flow to a depleted water region or vapor layer. Such arguments were supported by the microbubbles observed near hydrophobic surfaces [12], and flow models considering the thickness of the depleted water region or air gap were established [13]. Martini et al. [14] attributed the liquid slip to localized liquid atoms hopping at low levels of forcing, and to instantaneous motion of all liquid atoms at high levels of forcing. Nevertheless, the physical mechanism of fluid slip (damping) remains poorly understood.

Richardson [15] argued that, when the intermolecular attraction between liquid molecules and solid walls becomes much weaker than that between liquid molecules themselves, the liquid cannot fully wet the solid, so slip flow occurs over the solid surface. Compared with wettability, the interaction force between liquid molecules and a solid surface is a more intrinsic index of hydrophobicity of a solid surface, because it recognizes the existence of an adsorbed liquid layer, leading to negative slip lengths [16] and the fact that the liquid can slip on hydrophilic surfaces [17]. Microscopic molecular dynamics (MD) simulations [18, 19] and the mesoscopic lattice Boltzmann method (LBM) [20–26] explore the relationship between the slip magnitude and the fluid-solid (F-S) interaction strength.

The F-S interaction force is responsible for an apparent hydrophobic slip and can thus be quantified by the difference between hydrodynamic force on the hydrophilic surface as a no-slip boundary and on the hydrophobic surface with a slip flow [27, 28]. Moreover, it was recognized as an apparent extra attraction by Vinogradova [29, 30] and is determined by the total hydrodynamic force minus the van der Waals attraction, hydrophobic attraction, and Reynolds drag. The F-S interaction force is a few orders of magnitude greater than the van der Waals attraction and decays with the distance between liquid molecules and solid walls; however, it still cannot be expressed theoretically. The hydrophobic attraction force between solid surfaces [29, 30], which decays exponentially and by a power law, may provide a useful analogy for the F-S interaction force.

The F-S interaction was quantified in an MD simulation by using the Lennard-Jones potential with tunable interaction energies and molecular diameters to recover the contact angle. The LBM method used an exponentially decaying force or an interaction potential within a single lattice layer to model the F-S interaction. Assuming a no-slip boundary (no true slip on the solid surface), the interaction strength (and the decay length) was adjustable, which resulted in apparent fluid slip (damping) behavior. A continuous force was established between fluid particles and flat walls or at the particle level. In particular, the latter force function was used to investigate the liquid slip flow in porous media. Other LBM models were used to study the fluid slip on solid surfaces by combining the bounce-back and specular reflection boundary conditions [31, 32], linking the LBM relaxation time with the Kn number [33, 34] or using empirical constitutive relationships between slip velocity and Kn number [35, 36]. The mesoscopic LBM approach is advantageous because it considers the physics of the F-S interaction, which is impossible in continuum fluid dynamics. Also, it permits the simulation of slip-flow experiments on microfluidic devices and microporous media with acceptable computational costs compared with microscopic MD simulations.

Therefore, the present study exploits the LBM model framework to investigate the liquid slip flow between two confined parallel plates. By analogy with the hydrophobic attraction between solid surfaces [29, 30], we propose to use exponentially and power-law decaying force functions between fluid particles and two flat walls to describe the physics of fluids near solid surfaces. The details of the LBM model and the proposed continuous force functions are presented in the next section. In Sec. 3, the analytical solutions for density profile, velocity profile, slip length, and permeability ratio are derived based on the two continuous force functions, from which the dimensionless numbers for the corresponding slip-flow systems can be extracted. The analytical solutions thus obtained are closely consistent with the numerical solutions of an advanced LBM scheme, as presented in Sec. 4. Some useful implications in engineering contexts are discussed, and the permeability ratios are found to be simply determined by the dimensionless numbers as the fitted relationship formulated. In addition, we apply the analytical solutions to a benchmark slip-flow experiment. Finally, some concluding remarks are given in the last section.

## 2. Fluid-solid interaction in lattice Boltzmann model

### 2.1. Framework of lattice Boltzmann model

The LBM model is based on mesoscopic kinetic theory and has been widely used for simulations of immiscible two-phase flow [37, 38] and liquid slip flow. Here, the intermolecular interaction between liquid molecules is assumed to be negligible compared with the F-S interaction because, in the interior of the bulk fluid, the attraction forces from the neighboring liquid molecules can almost be offset for a liquid molecule. The thermodynamically stable bulk fluid is described by the single-phase LBM model. The particle distribution functions $f_k$ in the direction of the k-th discrete velocity obeys the Bhatnagar–Gross–Krook (BGK) lattice Boltzmann equation [39]

$$f_k(\mathbf{x} + \mathbf{e}_k \Delta t, t + \Delta t) - f_k(\mathbf{x}, t) = \frac{\Delta t}{\tau}\big[f_k^{eq}(\mathbf{x}, t) - f_k(\mathbf{x}, t)\big], \qquad (1)$$

where $\mathbf{x}, t$ are position and time, respectively, $\Delta t = 1$ is the time step, $\mathbf{e}_k$ is the discrete velocity from Ref. [26] for $D_2Q_9$ (nine velocities in two-dimensional space) and $D_3Q_{19}$ (nineteen velocities in three-dimensional space) LBMs, and $f_k^{eq}(\mathbf{x}, t)$ is the Maxwell–Boltzmann equilibrium distribution function as formulated in Ref. [26]. To recover the macroscopic Navier–Stokes (NS) equation, the relaxation time $\tau$ is related to the kinematic viscosity $\nu$ by $\nu = c_s^2 \Delta t (\tau/\Delta t - 0.5)$. In this isothermal LBM framework, the speed of sound $c_s$ is fixed at $c_s^2 = 1/3$. The left- and right-hand sides of Eq. (1) are respectively the streaming process passing the particle distribution to the neighboring nodes and the BGK collision operator partially relaxing the particle distribution to the equilibrium distribution.

The explicit force scheme (EFS) is used to quantify the momentum transfer when the external body force $\mathbf{F}$ is involved. The forcing term $\varphi_k$, which accounts for changes in the distribution function, and the transformation function $h_k$ are defined as [26, 40]

$$\varphi_k(\mathbf{x}, t) = \frac{\mathbf{F} \cdot (\mathbf{e}_k - \mathbf{u}^{eq})}{\rho c_s^2} f_k^{eq}(\mathbf{x}, t), \quad h_k(\mathbf{x}, t) = f_k(\mathbf{x}, t) - \frac{\Delta t}{2} \varphi_k(\mathbf{x}, t), \qquad (2)$$

where $\rho$ is the macroscopic fluid density, and $\mathbf{u}^{eq}$ is the equilibrium velocity equal to the macroscopic flow velocity $\mathbf{u}$. By directly incorporating $\varphi_k$ into the lattice Boltzmann equation (1) with the collision operator unchanged and applying $h_k$ to the resulting implicit formula, we obtain the following explicit expression of the new particle distribution function $h_k$:

$$h_k(\mathbf{x} + \boldsymbol{e}_k \Delta t, t + \Delta t) - h_k(\mathbf{x}, t) = \frac{1}{\tau}\left[f_k^{eq}(\mathbf{x}, t) - h_k(\mathbf{x}, t) - \frac{\Delta t}{2}\varphi_k(\mathbf{x}, t)\right] + \Delta t \varphi_k(\mathbf{x}, t). \tag{3}$$

The fluid density $\rho$ and the flow velocity $\mathbf{u}$ are then given by

$$\rho = \sum_k h_k, \quad \mathbf{u} = \frac{1}{\rho}\left(\sum_k h_k \mathbf{e}_k + \frac{\Delta t}{2}\mathbf{F}\right). \tag{4}$$

By using the equation of state of an ideal gas, the macroscopic fluid pressure $p^*$ is

$$p^* = c_s^2 \rho. \tag{5}$$

The EFS-LBM scheme can inhibit, to a great extent, the undesirable spurious velocities and the dependence of numerical results on the relaxation time. The LBM simulation of the steady-state flow runs until reaching the equilibrium criterion for flow velocity.

## 2.2. Fluid-solid interaction force

Combined with the bounce-back boundary condition [39] admitting no true slip on a solid surface, the F-S interaction force is the focus of this research and is responsible for the apparent fluid slip. Although a theoretical formulation of the F-S interaction force remains unavailable, it can be deduced by analogy with the hydrophobic attraction force between solid surfaces. Recognizing the nature of the F-S interaction force at the particle level [26] and its decay as an exponential function (EXF) or a power-law function (PLF) [29, 30], the F-S interaction force between fluid particles and flat walls can be obtained by integrating the local forces over the entire solid-wall surface. The forms of the EXF or PLF functions remain the same after integration. Thus, for a confined parallel-plate microchannel [see Fig. 1], we propose the following expression for the individual F-S interaction force $f_s$ between fluid particles and each wall:

$$f_s(y) = -\rho(y)g_s \exp\left(-\frac{y}{\eta}\right), \tag{6a}$$

$$f_s(y) = -\frac{\rho(y)g_s}{y^n}, \tag{6b}$$

where $y$ is the spanwise position of fluid particles in the given coordinate system, $g_s$ is the interaction strength, which determines the magnitude of interaction force, and $\eta, n$ are the decay length and the decay rate, respectively, that control the decay of the two force curves. Considering that the gap between two parallel plates is very narrow, the effect of the F-S interaction force exerted by the other wall should be included. Thus, the net interaction force $F_s$ from the two flat walls separated by a $2R$ gap can be expressed by

$$F_s(y) = -\rho(y)g_s\left[\exp\left(-\frac{y}{\eta}\right) - \exp\left(-\frac{2R-y}{\eta}\right)\right], \tag{7a}$$

$$F_s(y) = -\rho(y)g_s\left[\frac{1}{y^n} - \frac{1}{(2R-y)^n}\right]. \tag{7b}$$

The interaction force is applied to the fluid particles and so is related to the number density of fluid particles. In this way, the interaction force changes the fluid density and dynamic viscosity near the solid surface (assuming kinematic viscosity to be constant). The continuous EXF and PLF force functions in Eq. (7) allow the simulation of density and velocity profiles near the solid surface. Moreover, the excessive pressure due to the F-S interaction potential plus the ideal gas pressure contribute to the macroscopic fluid pressure $p$, as formulated by [22, 23]

$$p = c_s^2 \rho + \int_0^y F_s(s) ds. \tag{8}$$

Substituting Eq. (7) into Eq. (8) gives the explicit formulations for macroscopic fluid pressure with the EXF and PLF force functions:

$$p(y) = c_s^2 \rho(y) - g_s \int_0^y \rho(s) \left[ \exp\left(-\frac{s}{\eta}\right) - \exp\left(-\frac{2R-s}{\eta}\right) \right] ds, \tag{9a}$$

$$p(y) = c_s^2 \rho(y) - g_s \int_0^y \rho(s) \left[ \frac{1}{s^n} - \frac{1}{(2R-s)^n} \right] ds. \tag{9b}$$

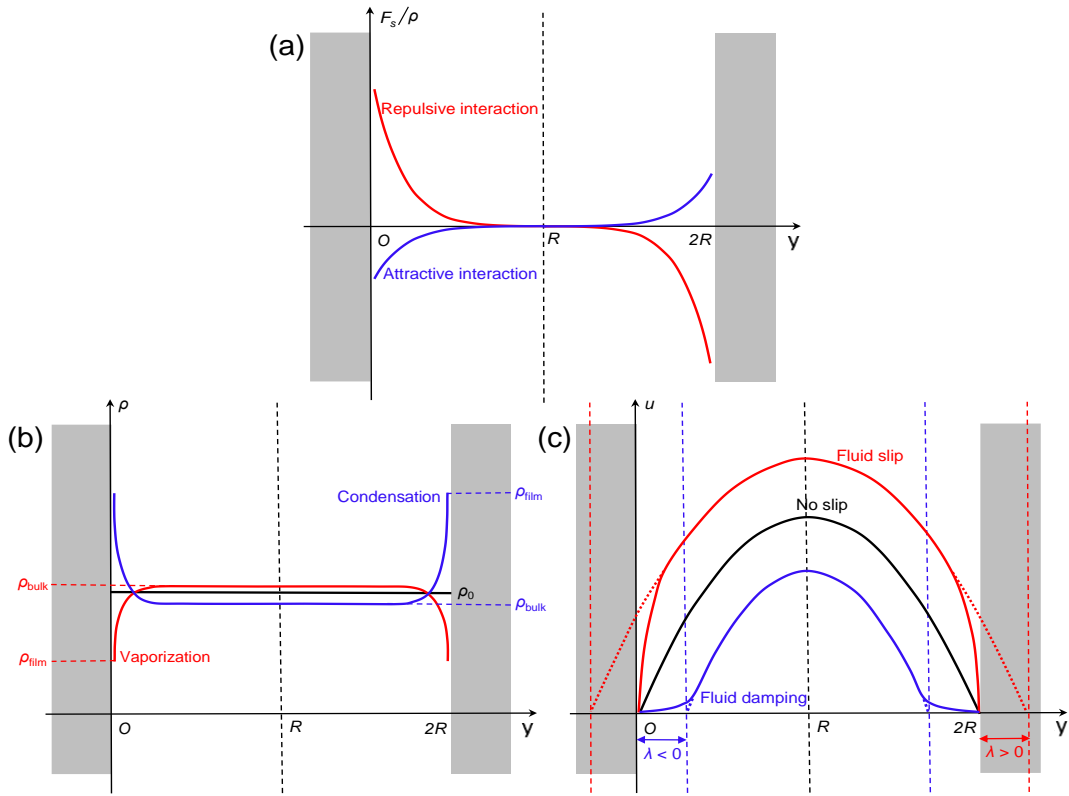

**Fig. 1.** Diagram of a confined parallel-plate microchannel with a gap diameter of $2R$ and an upward flow. The microchannel may be two dimensional (with two straight walls) or three dimensional (with two flat surfaces). (a) The force curves of the net interaction from the two flat walls for the repulsive and attractive F-S interactions. (b) The density profiles $\rho(y)$ and the densities $\rho_{\text{film}}$ of the film liquid and $\rho_{\text{bulk}}$ of the bulk liquid during the vaporization and condensation effects. (c) The velocity profiles $u(y)$ and the slip lengths $\lambda$ for different flow behaviors (fluid slip, fluid damping and no slip).

Figure 1(a) shows the decay profiles of the two force functions. The magnitude of the net interaction force reaches a maximum near the solid surface and decays to zero in the middle of the channel, where the individual interaction forces from the two flat walls are entirely offset. The net (and even individual) F-S interaction force is repulsive (attractive) for $g_s > 0$ ($g_s < 0$) and decays faster for smaller $\eta$ (larger $n$). Both larger $|g_s|$ and longer interaction distance (slower decaying process) indicate stronger F-S interactions. However, we should avoid the extreme case, where the parallel-plate microchannel is very confined (relative to the given individually decaying interaction force), and the net interaction force accordingly becomes very weak, even for the strong individual interaction force. This leads to the physically inconsistent result that the effect of the net F-S interaction on the macroscale flow properties decreases as the gap narrows between the two flat walls, and our mesoscopic LBM model fails. This extreme case is characterized by a high Kn number (Kn > 0.1), where the collision between liquid molecules themselves becomes much less frequent, and the NS equation cannot adequately describe the macroscopic fluid flow. In this case, the intermolecular interaction between liquid molecules themselves and the F-S interaction is better described by the Lennard–Jones potential in the microscopic MD simulation.

### 2.3. Effect of fluid-solid interaction on flow properties

Depending on their physicochemical properties, the solid walls tend to exert a repulsive (or attractive) force on the fluid particles at the mesoscale. The liquid molecules, as a result, are expelled from (or gather near) solid surfaces, and the liquid density in a very thin layer close to solid surfaces decreases (or increases) slightly with respect to the bulk liquid density, which is referred to as the vaporization (or condensation) effect [see Fig. 1(b)]. The difference in density between the film liquid density and the free liquid density has been confirmed experimentally on bound water adsorbed on various surfaces [41–46]. The bound-water density increases with its reduced content (film thickness) for the attractive interaction, showing that the film liquid density rises as the liquid layer thins (closer to the solid surface). The ratio of film liquid density to bulk liquid density is

$$\rho_{ra} = \frac{\rho_{film}}{\rho_{bulk}}. \tag{10}$$

In the analytical solutions obtained in the next section, the fluid densities at $y_1 = 1$ lattice unit (lu) (closest to the solid surface in the lattice grid system) and $y_2 = R$ are used to represent the film and bulk fluid densities; that is, $\rho_{film} = \rho(1)$ and $\rho_{bulk} = \rho(R)$. Table 1 summarizes the ranges of density ratio for various experimental conditions.

**Table 1** Range of density ratio observed in various experimental investigations.

| Ref. | Liquid | Solid material | $\rho_{ra}$ range |
|------|--------|----------------|-------------------|
| [41] | Water | Silica gel | 0.54–1.03 |
| [42] | Water | Bentonite clay | 0.97–0.98 |
| [43] | Water | Kaolinite | 0.99–1.68 |
| | | Montmorillonite | 1.02–1.41 |
| [44] | Water | Glukhovtes kaolin | 1.1–1.5 |
| | | Cherkassy bentonite | 1.1–1.8 |
| [45] | Water | Natural kaolinite | 0.98–1.22 |
| | | Natural montmorillonite | 1.06–1.15 |
| [46] | Water | Cellulose | 0.2–1.3 |

The vaporized (condensed) film liquid leads to a reduction (increase) of shear viscosity near the solid surface, and the resulting flow becomes faster (slower) near the solid surface and in the bulk fluid than that under the no-slip condition, which is the so-called apparent fluid slip (damping) behavior. This phenomenon is consistent with the argument [10, 11] that the slip flow is due to (and occurs on) the depleted water region or vapor layer near the hydrophobic surface. Under the no-slip condition, the velocity $u^*$ profile between two parallel plates is a Poiseuille-type curve [see Fig. 1(c)] and is given by

$$u^*(y) = \frac{g}{2\nu} y(2R - y), \tag{11}$$

where $g$ is the acceleration due to the external body force to drive the fluid flow. In the presence of a repulsive (attractive) F-S interaction, the near-wall velocity still vanishes at the wall, and the bulk velocity remains a Poiseuille-type curve but extrapolates to zero outside (inside) the solid wall [26]. By introducing a slip length $\lambda$ to describe the apparent slip (damping) behavior, the bulk velocity $u$ can be approximated as follows:

$$u(y) = \frac{g}{2\nu'} (y + \lambda)(2R + \lambda - y). \tag{12}$$

The F-S interaction cannot affect the kinematic viscosity $\nu'$ of the bulk fluid [26], such as $\nu' = \nu$. The role of the F-S interaction lies in its effect on the fluid density near the solid surface. In this way, the F-S interaction can be coupled with the change of dynamic viscosity and the resulting slip flow. Equation (12) overestimates (underestimates) the near-wall velocity and the volumetric flux when $\lambda > 0$ ($\lambda < 0$). The F-S interaction effect on the bulk flow property is equivalent to the gap between the two flat walls increasing to $(2R + 2\lambda)$. The slip lengths of typical and engineered hydrophobic surfaces range from 1–100 nm [1] to around 1 micron (μm) [5].

The permeability ratio $k_{\mathrm{ra}}$ (called the "flow-enhancement ratio") is the ratio of the flux $Q_{\mathrm{slip}}$ in the slip flow to the flux $Q_{\mathrm{no-slip}}$ in the no-slip flow and describes the change in flow property:

$$k_{\mathrm{ra}} = \frac{Q_{\mathrm{slip}}}{Q_{\mathrm{no-slip}}}, \tag{13}$$

where $Q_{\mathrm{no-slip}}$ and $Q_{\mathrm{slip}}$ are the volumetric flux or mass flux. The flow-enhancement ratio of thin channels varies from 1.05 [3] to 1.30 [4]. For the fluid slip (damping) behavior where $g_{\mathrm{s}} > 0$ ($g_{\mathrm{s}} < 0$), the fictitious no-slip plane lies within the solid (fluid), resulting in a positive (negative) slip length, given by $\lambda > 0$ ($\lambda < 0$), and the permeability of the microchannel increases (decreases) with $k_{\mathrm{ra}} > 1$ ($k_{\mathrm{ra}} < 1$).

## 3. Derivations of analytical solutions

In the hydrodynamic limit of small Kn number (Kn < 0.1), the lattice Boltzmann equation (1) becomes the NS equation in the continuum fluid dynamics by using the Chapman–Enskog expansion [22, 23, 39]:

$$\partial_t(\rho u_i) + (u_j \partial_j)(\rho u_i) = -\partial_i p + F_i + \nu \partial_j(\rho \partial_j u_i + \rho \partial_i u_j), \tag{14a}$$

$$\partial_t \rho + \partial_i(\rho u_i) = 0, \tag{14b}$$

where $u_i$ and $F_i$ are the flow velocity and the total force in direction i. For the liquid slip flow between the two confined parallel plates in Fig. 1, i (j) represents the streamwise $x$ or spanwise $y$ direction. Considering the net F-S interaction force from the two solid walls plus an external body force driving the liquid flow, a total force of $\mathbf{F} = \mathbf{F}_s + \rho \mathbf{g}$ (where $\mathbf{F}$, $\mathbf{F}_s$, and $\mathbf{g}$ are the vector form of $F_i$, $F_s$, and $g$), arising in the EFS-LBM scheme in Eq. (2), is exerted on the fluid.

When the liquid flow reaches the steady state and maintains a laminar regime, $\partial_t \rho = 0$, $\partial_t(\rho u_i) = 0$, and the inertial term in the momentum balance equation is neglected. The lateral flow is stagnant, but the fluid density may vary with the distance from the solid wall due to the F-S interaction force. No external body force is present and only the pressure-gradient term remains. Thus, Eq. (14a) in the $y$ direction can be rewritten as

$$\partial_y p = 0. \tag{15}$$

For a flow channel with homogeneous gap size, the flow velocity and the fluid density remain constant in the streamwise direction. No pressure gradient exists and only an external body force is applied. Thus, Eq. (14a) in the $x$ direction reduces to

$$\rho(y)g + \nu \partial_y[\rho(y)\partial_y u] = 0. \tag{16}$$

The continuity equation is satisfied for a finite volume of fluid. Integrating Eq. (14b) over the gap leads to $\partial_t[\int_0^R \rho(y)dy] = 0$, which means that $\int_0^R \rho(y)dy$ remains constant over time. With the specified initial homogeneous fluid density $\rho_0$, the macroscale mass balance equation is

$$\int_0^R \rho(y)dy = \rho_0 R. \tag{17}$$

Additionally, the no-slip boundary condition at solid surface and the maximum velocity in the middle of the gap can be both applied,

$$u(0) = 0, \partial_y u(R) = 0. \tag{18}$$

From the governing equations and conditions imposed by Eqs. (15)–(18), we can derive analytical solutions for the density profile, velocity profile, slip length, and permeability ratio for the liquid slip flow between the two confined parallel plates shown in Fig. 1.

### 3.1. Density and velocity profiles

Replacing the macroscale fluid pressure in Eq. (15) with Eq. (9), considering the mesoscale F-S interaction force, gives

$$\partial_y \rho - \frac{g_s}{c_s^2} \rho(y) \left[ \exp\left(-\frac{y}{\eta}\right) - \exp\left(-\frac{2R-y}{\eta}\right) \right] = 0, \tag{19a}$$

$$\partial_y \rho - \frac{g_s}{c_s^2} \rho(y) \left[ \frac{1}{y^n} - \frac{1}{(2R-y)^n} \right] = 0. \tag{19b}$$

Set $a = g_s/c_s^2$, which refers to the interaction strength in the following unless stated otherwise. Equation (19) can be rewritten as

$$\frac{\partial_y \rho}{\rho(y)} = a \left[ \exp\left(-\frac{y}{\eta}\right) - \exp\left(-\frac{2R-y}{\eta}\right) \right], \tag{20a}$$

$$\frac{\partial_y \rho}{\rho(y)} = a \left[ \frac{1}{y^n} - \frac{1}{(2R-y)^n} \right]. \tag{20b}$$

By solving the homogeneous ordinary differential equations (20) for the fluid density with the aid of the auxiliary condition (17), the analytical solution for the density profile based on EXF and PLF force functions are

$$\rho(y) = \rho_0 A_1 \exp[-aw_1(y)], \tag{21a}$$

$$\rho(y) = \rho_0 A_2 \exp[-aw_2(y)], \tag{21b}$$

where

$$w_1(y) = \eta \left[ \exp\left(-\frac{y}{\eta}\right) + \exp\left(-\frac{2R-y}{\eta}\right) \right], \quad A_1 = \frac{R}{\int_0^R \exp[-aw_1(y)] \mathrm{d}y},$$

$$w_2(y) = \frac{1}{(n-1)} \left[ \frac{1}{y^{n-1}} + \frac{1}{(2R-y)^{n-1}} \right], \quad A_2 = \frac{R}{\int_0^R \exp[-aw_2(y)] \mathrm{d}y}.$$

By rearranging Eq. (16), we obtain

$$\partial_y^2 u + \frac{\partial_y \rho}{\rho(y)} \partial_y u = -b, \tag{22}$$

where in the following, unless stated otherwise, $b = g/\nu$ (with $\nu$ assumed to be constant) is the magnitude of the external body force. Equation (22) is the inhomogeneous ordinary differential equation for the flow velocity. By applying the condition (18), the analytical solutions for the velocity profile based on EXF and PLF force functions are

$$u(y) = b \int_0^y \int_r^R \exp\{a[w_1(r) - w_1(s)]\} \mathrm{d}s \, \mathrm{d}r, \tag{23a}$$

$$u(y) = b \int_0^y \int_r^R \exp\{a[w_2(r) - w_2(s)]\} \mathrm{d}s \, \mathrm{d}r. \tag{23b}$$

## 3.2. Slip length and permeability ratio

To obtain the slip length, Eq. (23) should be transformed into the same form as Eq. (12) for the bulk velocity profile. We reformulate the velocity profiles for the two force functions as

$$u(y) = \frac{b}{2} q_{\mathrm{m}}(y) \ (m = 1, 2), \tag{24}$$

where

$$q_1(y) = 2 \int_0^y \int_r^R \exp\{a[w_1(r) - w_1(s)]\} \mathrm{d}s \, \mathrm{d}r,$$

$$q_2(y) = 2 \int_0^y \int_r^R \exp\{a[w_2(r) - w_2(s)]\} \mathrm{d}s \, \mathrm{d}r.$$

The expression for $q_{\mathrm{m}}(y)$ can be approximated by a second-order Taylor series expansion around the middle of the gap ($y \approx R$). Equation (24) may then be rearranged as

$$u(y)|_{y \approx R} \cong \frac{b}{2} \left[ \frac{q_{\mathrm{m}}(R)}{0!} + \frac{\partial_y[q_{\mathrm{m}}(R)]}{1!}(y-R) + \frac{\partial_y^2[q_{\mathrm{m}}(R)]}{2!}(y-R)^2 \right] \ (m = 1, 2). \tag{25}$$

The function $q_{\mathrm{m}}(x)$ and its derivatives at $x = R$ are

$$q_{\mathrm{m}}(R) = q_{\mathrm{m}}(R), \ \partial_y[q_{\mathrm{m}}(R)] = 0, \ \partial_y^2[q_{\mathrm{m}}(R)] = -2 \ (m = 1, 2). \tag{26}$$

Substituting the computed Taylor coefficients into Eq. (25) produces

$$u(y)|_{y \approx R} \cong \frac{b}{2} [-y^2 + 2Ry + q_{\mathrm{m}}(R) - R^2] \ (m = 1, 2). \tag{27}$$

Although Eq. (27) approximates Eq. (24) only at $y \approx R$, it represents the bulk velocity profile in a manner equivalent to Eq. (12) because it retains the Poiseuille-type nature and the same kinematic viscosity. By comparing Eq. (27) with Eq. (12), we obtain a parabolic equation with the unknown slip length:

$$\lambda^2 + 2R\lambda + R^2 - q_{\mathrm{m}}(R) = 0 \ \ (\mathrm{m} = 1, 2). \tag{28}$$

The analytical solution for the slip length (normalized by the gap between the two flat walls) is then derived as

$$\frac{\lambda}{R} = \frac{1}{R}\sqrt{q_{\mathrm{m}}(R)} - 1 \ \ (\mathrm{m} = 1, 2). \tag{29}$$

Substituting the expressions for $q_{\mathrm{m}}(R)$ into Eq. (29) gives respectively the analytical normalized slip length based on the EXF and PLF force functions

$$\frac{\lambda}{R} = \frac{1}{R}\sqrt{2\int_0^R \int_r^R \exp\{a[w_1(r) - w_1(s)]\}\mathrm{d}s\,\mathrm{d}r} - 1, \tag{30a}$$

$$\frac{\lambda}{R} = \frac{1}{R}\sqrt{2\int_0^R \int_r^R \exp\{a[w_2(r) - w_2(s)]\}\mathrm{d}s\,\mathrm{d}r} - 1. \tag{30b}$$

The volumetric flux and mass flux can be calculated based on the analytical results for density and velocity. As given by Eq. (13), the volumetric permeability ratio $k_{\mathrm{rav}}$ and the mass permeability ratio $k_{\mathrm{ram}}$ are expressed by

$$k_{\mathrm{rav}} = \frac{\int_0^R u(y)\mathrm{d}y}{\int_0^R u^*(y)\mathrm{d}y}; \ k_{\mathrm{ram}} = \frac{\int_0^R \rho(y)u(y)\mathrm{d}y}{\int_0^R \rho_0 u^*(y)\mathrm{d}y}. \tag{31}$$

Inserting the Poiseuille-type velocity from Eq. (11) and the slip velocity from Eq. (23) into Eqs. (31) produces respectively the volumetric permeability ratios based on the EXF and PLF force functions:

$$k_{\mathrm{rav}} = \frac{3}{R^3}\int_0^R \int_0^y \int_r^R \exp\{a[w_1(r) - w_1(s)]\}\mathrm{d}s\,\mathrm{d}r\,\mathrm{d}y, \tag{32a}$$

$$k_{\mathrm{rav}} = \frac{3}{R^3}\int_0^R \int_0^y \int_r^R \exp\{a[w_2(r) - w_2(s)]\}\mathrm{d}s\,\mathrm{d}r\,\mathrm{d}y. \tag{32b}$$

Similarly, the mass permeability ratios based respectively on the EXF and PLF force functions are

$$k_{\mathrm{ram}} = \frac{3A_1}{R^3}\int_0^R \int_0^y \int_r^R \exp\{a[-w_1(y) + w_1(r) - w_1(s)]\}\mathrm{d}s\,\mathrm{d}r\,\mathrm{d}y, \tag{33a}$$

$$k_{\mathrm{ram}} = \frac{3A_2}{R^3}\int_0^R \int_0^y \int_r^R \exp\{a[-w_2(y) + w_2(r) - w_2(s)]\}\mathrm{d}s\,\mathrm{d}r\,\mathrm{d}y. \tag{33b}$$

The analytical solutions in Eqs. (21), (23), (30), (32), and (33) are derived based on the macroscopic fluid pressure equation (9) with the two continuous force functions and the macroscopic NS equation (14) recovered from the lattice Boltzmann equation. These equations directly relate the macroscale density and velocity, normalized slip length and volumetric and mass permeability ratios to mesoscale F-S interaction parameters and flow-channel gap size.

These analytical solutions contain single, double, and triple integrals, which can be solved numerically by using MATLAB. The EXF function is infinitely differentiable, so the analytical solutions based on the EXF force functions are numerically solvable for arbitrary values of $a, \eta$, and $R$. The PLF function has singularities at $y = 0$ $(2R)$ which remain in the analytical solutions. However, the analytical solutions based on the PLF force function can be solved numerically over a broad range of $a, n$, and $R$.

### 3.3. Dimensionless numbers

The nondimensional form of the analytical solutions can lead to the dimensionless numbers for the slip flow systems. Let us define

$$\rho' = \frac{\rho}{\rho_0}, \ u' = \frac{2u}{bR^2}, \ y' = \frac{y}{R}, \ r' = \frac{r}{R}, \ s' = \frac{s}{R}.$$

In the nondimensional form, the density and velocity from Eqs. (21) and (23) are

$$\rho'(y') = A_1 \exp[-a\eta w_1'(y')], \tag{34a}$$

$$\rho'(y') = A_2 \exp\left[-\frac{a}{R^{n-1}} w_2'(y')\right], \tag{34b}$$

$$u'(y') = 2 \int_0^{y'} \int_{r'}^1 \exp\{a\eta[w_1'(r') - w_1'(s')]\} \mathrm{d}s' \, \mathrm{d}r', \tag{35a}$$

$$u'(y') = 2 \int_0^{y'} \int_{r'}^1 \exp\left\{\frac{a}{R^{n-1}}[w_2'(r') - w_2'(s')]\right\} \mathrm{d}s' \, \mathrm{d}r'. \tag{35b}$$

where

$$w_1'(y') = \frac{w_1(y)}{\eta} = \exp\left(-\frac{y'}{\eta/R}\right) + \exp\left(-\frac{2-y'}{\eta/R}\right), \quad A_1 = \frac{1}{\int_0^1 \exp[-a\eta w_1'(y')] \mathrm{d}y'},$$

$$w_2'(y') = \frac{w_2(y)}{1/R^{n-1}} = \frac{1}{(n-1)}\left[\frac{1}{y'^{n-1}} + \frac{1}{(2-y')^{n-1}}\right], \quad A_2 = \frac{1}{\int_0^1 \exp\left[-\frac{a}{R^{n-1}} w_2'(y')\right] \mathrm{d}y'}.$$

These equations show that $w_1'(y')$ and $w_2'(y')$ are nondimensional and are functions of the dimensionless numbers $\eta/R$ and $(n-1)$, respectively. The ratio $\eta/R$ is the decay length normalized by the channel size for the EXF force function, and $(n-1)$ is the decay rate of the F-S interaction for the PLF force function. The extraction of $\eta$ and $1/R^{n-1}$ from $w_1(y)$ and $w_2(y)$ to obtain $w_1'(y')$ and $w_2'(y')$ leads to the new terms $a\eta$ and $a/R^{n-1}$ arising in the expressions for $A_1$ and $A_2$ and in the right-hand side of Eqs. (34) and (35), respectively. Through dimensional analysis based on Eq. (7), $a\eta = g_s\eta/c_s^2$ and $a/R^{n-1} = g_s/(R^{n-1}c_s^2)$ for the EXF and PLF force functions, respectively, are both nondimensional. Therefore, the nondimensional density and velocity profiles for EXF and PLF force functions are determined solely by the dimensionless numbers $(a\eta, \eta/R)$ and $(a/R^{n-1}, n-1)$, respectively. The same observation can also be made about the normalized slip length and the volumetric and mass permeability ratios, which take the form

$$\frac{\lambda}{R} = \sqrt{2 \int_0^1 \int_{r'}^1 \exp\{a\eta[w_1'(r') - w_1'(s')]\} \mathrm{d}s' \, \mathrm{d}r'} - 1, \tag{36a}$$

$$\frac{\lambda}{R} = \sqrt{2 \int_0^1 \int_{r'}^1 \exp\left\{\frac{a}{R^{n-1}}[w_2'(r') - w_2'(s')]\right\} \mathrm{d}s' \, \mathrm{d}r'} - 1, \tag{36b}$$

$$k_{\mathrm{rav}} = 3 \int_0^1 \int_0^{y'} \int_{r'}^1 \exp\{a\eta[w_1'(r') - w_1'(s')]\} \mathrm{d}s' \, \mathrm{d}r' \, \mathrm{d}y', \tag{37a}$$

$$k_{\mathrm{rav}} = 3 \int_0^1 \int_0^{y'} \int_{r'}^1 \exp\left\{\frac{a}{R^{n-1}}[w_2'(r') - w_2'(s')]\right\} \mathrm{d}s' \, \mathrm{d}r' \, \mathrm{d}y', \tag{37b}$$

$$k_{\mathrm{ram}} = 3A_1 \int_0^1 \int_0^{y'} \int_{r'}^1 \exp\{a\eta[-w_1'(y') + w_1'(r') - w_1'(s')]\} \mathrm{d}s' \, \mathrm{d}r' \, \mathrm{d}y', \tag{38a}$$

$$k_{\mathrm{ram}} = 3A_2 \int_0^1 \int_0^{y'} \int_{r'}^1 \exp\left\{\frac{a}{R^{n-1}}[-w_2'(y') + w_2'(r') - w_2'(s')]\right\} \mathrm{d}s' \, \mathrm{d}r' \, \mathrm{d}y'. \tag{38b}$$

As analyzed above, $(a\eta, \eta/R)$ and $(a/R^{n-1}, n-1)$ can be considered as the dimensionless numbers for slip flow systems based on the EXF and PLF force functions, respectively. $\eta/R$ and $(n-1)$ represent the range of F-S interaction near the solid surface, and $a\eta = g_s\eta/c_s^2$ and $a/R^{n-1} = g_s/(R^{n-1}c_s^2)$ provide comprehensive descriptions of the interaction strength. Large $g_s$, large $\eta$, and small $R$ (i.e., a narrow channel) can strengthen the effect of F-S interaction. Although $c_s^2$ in Eq. (5) is fixed at $1/3$ in this isothermal LBM framework, it can be used to measure the compressibility of the specific fluid, with the fluid being more easily compressed with a higher value of $1/c_s^2 = \Delta\rho/\Delta p$. The larger density gradient due to the F-S interaction may result in a more significant slip-flow effect.

## 4. Results and discussions

### 4.1. Validation of analytical profiles of density and velocity

Each of the analytical solutions derived based on the EXF and PLF force functions has two free interaction parameters. Figure 2 shows the density ratio with respect to these interaction parameters. With a larger $|a|$ and a larger $\eta$ (smaller $n$), which indicates a stronger interaction and a more extended interaction range, the density ratio deviates further from unity, implying more vaporization (or condensation). The density ratio can even reach as high as $10^4$ and as low as $10^{-8}$.

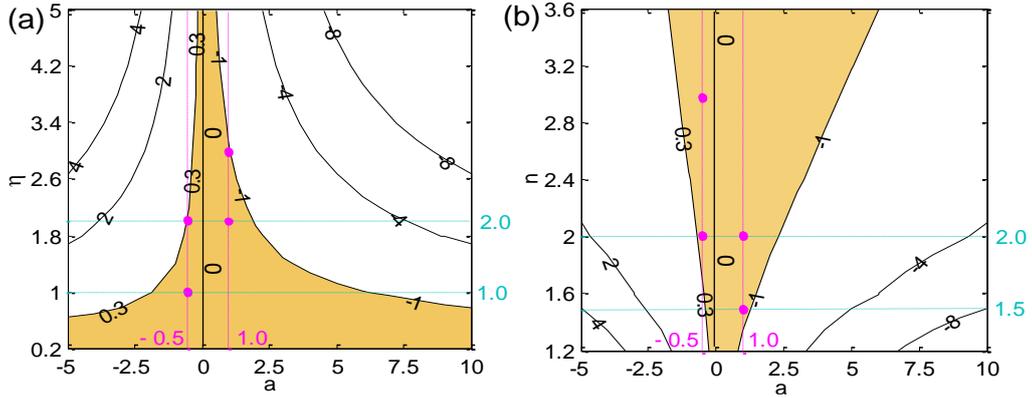

**Fig. 2.** The density ratio $\log_{10}(\rho_{ra})$ (curves labeled by values) with respect to (a) the interaction parameters $a$ and $\eta$ for the EXF force function and (b) $a$ and $n$ for the PLF force functions, for a gap size $R = 100$ lu. The shaded region is the range of (a) $a$ and $\eta$ and of (b) $a$ and $n$ where $0.1 < \rho_{ra} < 2.0$, as observed in various experimental investigations [41–46]. The eight solid circles show the values of (a) $a$ and $\eta$ and of (b) $a$ and $n$ used for comparing analytical solutions with LBM solutions. $\eta = 1.0,\ 2.0$ and $n = 1.5,\ 2.0$ are used for investigating the analytical solutions that depend on $a$. All the variables are presented in lattice units (the same holds for the following figures unless stated otherwise).

However, the ranges of $a$ and $\eta$ or $n$ are not arbitrary and should be physically consistent with the observed density ratio of film fluid to bulk fluid, as summarized in Table I. Specifically,

$$0.1 < \rho_{ra} < 2.0. \tag{39}$$

The shaded region in Fig. 2 shows the ranges of $a$ and $\eta$ $(n)$ for the EXF (PLF) force function corresponding to the experimentally observed range of density ratio. Following this guideline, the reasonable range of $a$ becomes very limited for the large $\eta$ (small $n$) and very broad for the small $\eta$ (large $n$).

Figure 3 plots the analytical solutions for density [Eq. (21)] and velocity [Eq. (23)] using the values of $a$ and $\eta$ $(n)$ taken from the shaded range in Fig. 2. For the same $|a|$ but a larger $\eta$ (smaller $n$), the film and bulk fluid densities are further away from the initial density, and the maximum velocities deviate more significantly from that in the no-slip flow. Note that the bulk fluid density based on the PLF force function is not as flat as that based on the EXF force function.

The analytical solutions based on the two force functions are in excellent agreement with the numerical solutions using the EFS-LBM scheme, as shown in Fig. 3, which validates the analytical solutions solved numerically by MATLAB. These analytical solutions can thus provide benchmark solutions of liquid slip flow for various advanced LBM schemes and experimental investigations.

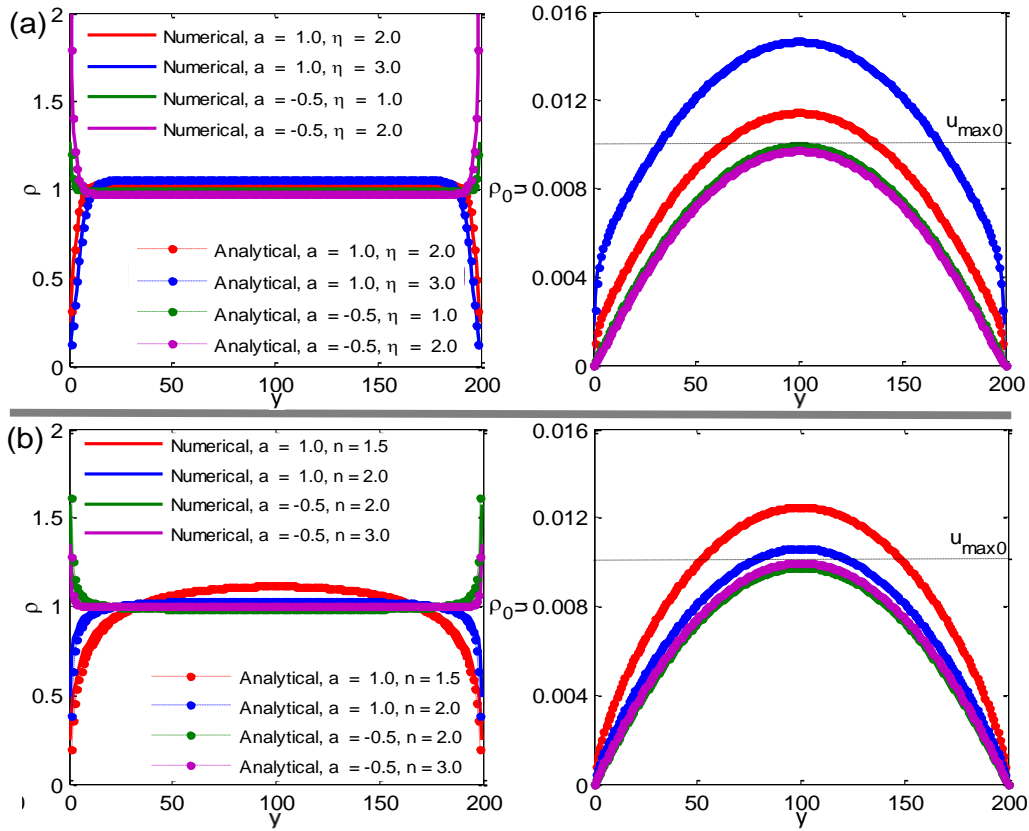

**Fig. 3.** Comparison of numerical and analytical solutions for density and velocity based on the (a) EXF and (b) PLF force functions, respectively. $\rho_0$ is the specified homogeneous fluid density at the initial time, and $u_{max0}$ is the maximum velocity in the middle of the gap under the no-slip boundary condition.

## 4.2. Implications of analytical slip length and permeability ratio

In engineering practice, such as for water in a tight sandstone reservoir or oil in a shale matrix, the slip length and the permeability ratio are always used to quantify the degree of liquid slip flow through tight porous media, which is controlled by the physicochemical property of the solid surface and the size of the gap in the flow channel. As shown in Fig. 4, within the reasonable range of interaction strength, the normalized slip length in Eq. (30) and the volumetric and mass permeability ratios in Eqs. (32) and (33) increase superlinearly with $a$ for $\eta = 1.0,\ 2.0$ and $n = 1.5,\ 2.0$, respectively. In addition, the curves for normalized slip length and volumetric and mass permeability ratios are steeper for $\eta = 2.0$ ($n = 1.5$) than those for $\eta = 1.0$ ($n = 2.0$), implying a more significant F-S interaction effect on the flow properties. The analytical volumetric and mass permeability ratios based on the two force functions cover the measured range of flow-enhancement ratio [3, 4].

Two other interesting results are obtained from Fig. 4. 1) Within the physically consistent range of density ratio, the normalized slip length and the volumetric and mass permeability ratios based on the EXF force function cover a broader range than those based on the PLF force function. 2) The curves for the volumetric and mass permeability ratios based on the EXF force function almost collapse, whereas that for the mass permeability ratio based on the PLF force function are a bit steeper than the curve for the volumetric permeability ratio. These two results are manifested in the different behavior of film and bulk fluid density based on the two force functions.

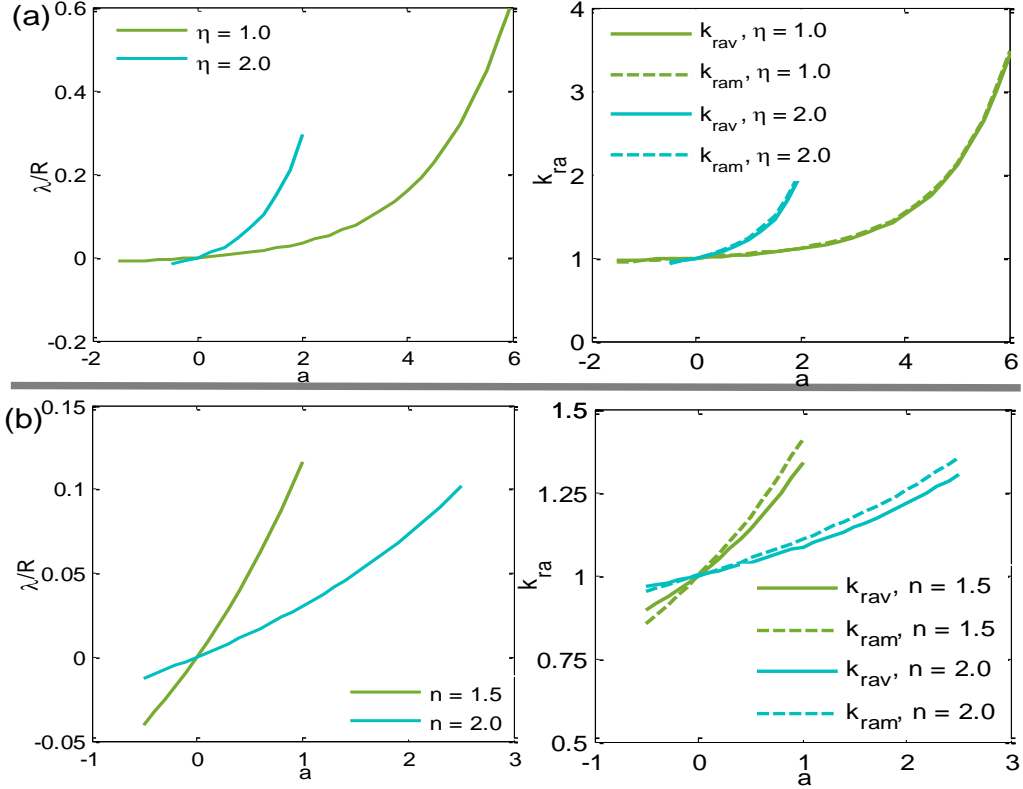

**Fig. 4.** Normalized slip length and permeability ratio as functions of interaction strength based on (a) EXF and (b) PLF force functions, respectively, for the gap size of $R = 100$ lu.

The slip length and the volumetric permeability ratio can be directly correlated on the macroscale. Neglecting the effect of near-wall velocity on the volumetric flux and inserting the no-slip velocity profile [Eq. (11)] and the approximate bulk velocity [Eq. (12)] into Eq. (31), we can estimate the volumetric permeability ratio as follows [26]:

$$k_{\mathrm{rav}} \approx \begin{cases} 1 + 3\left(\frac{\lambda}{R} + \frac{\lambda^2}{2R^2}\right), & \lambda > 0, \\ \left(1 + \frac{\lambda}{R}\right)^3, & \lambda \leq 0. \end{cases} \tag{40}$$

Figure 5 shows the estimated volumetric permeability ratio from Eq. (40) as a function of the normalized slip length, which illustrates that the two variables are positively correlated.

Reorganizing the analytical results shown in Fig. 4 gives the analytical volumetric permeability ratio as a function of normalized slip length. The estimated volumetric permeability ratios based on the EXF force function is highly consistent with the analytical result, whereas the two results for the PLF force function tend to diverge as $|\lambda|/R$ increases (the F-S interaction strengthens). These results reflect the different near-wall velocity profiles resulting from the two force functions.

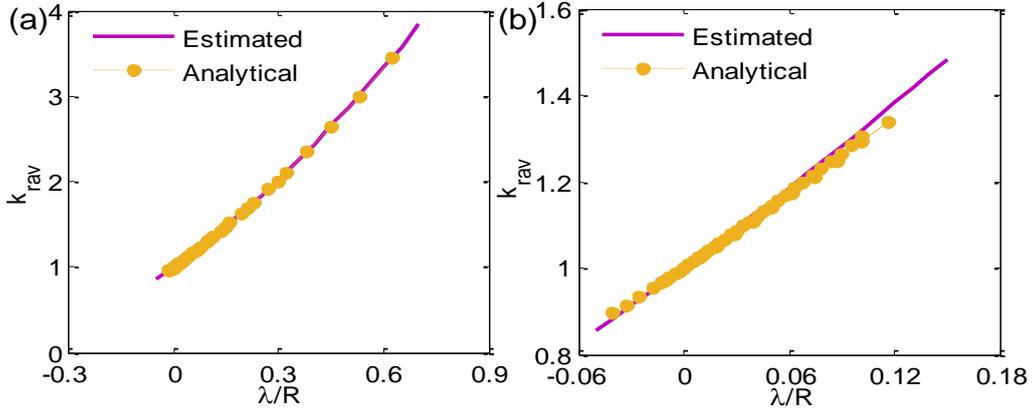

**Fig. 5.** Estimated volumetric permeability ratio from Eq. (40) compared with the analytical volumetric permeability ratios from Eq. (32) as a function of the normalized slip length from Eq. (30) based on the (a) EXF and (b) PLF force functions.

As presented in Fig. 6, the normalized slip length $|\lambda|/R$ (evaluating how the slip-flow affects the macroscopic flow) based on the two force functions increases rapidly with decreasing gap in the parallel-plate channel for each pair $(a, \eta)$ and $(a, n)$. Specifically,

$$\frac{\mathrm{d}(|\lambda|/R)}{\mathrm{d}R} < 0. \tag{41}$$

Also, the volumetric and mass permeability ratios increase (decrease) nonlinearly as the gap narrows for $a > 0$ ($a < 0$). These results are consistent with the observation that the F-S interaction (liquid slip flow) is quite significant in tight porous media.

Note that, in Fig. 6, the difference between the volumetric permeability ratio and the mass permeability ratio based on the PLF force function is greater than that based on the EXF force function. As the gap increases, the normalized slip length and the permeability ratios based on the EXF force function converge rapidly to zero and unity, respectively, whereas the normalized slip length and the permeability ratios based on the PLF force function converge much more slowly to zero and unity, respectively. These results indicate that the profiles of the bulk fluid density and the flow velocity differ for the two force functions.

When the gap is narrower than a critical gap $R_0$, as given in Fig. 7, the normalized slip length $|\lambda|/R$ based on the EXF force function decreases rapidly as the gap narrows for arbitrary values of $\eta$, and a similar trend of $|\lambda|/R$ based on the PLF force function is also found for $n \geq 1.2$. These results imply a weakened liquid slip flow in tight porous media with a narrow gap and are inconsistent with physical observations [8, 9]. As discussed in Sec. 2.2, the extremely narrow gap results in a high Kn number (Kn > 0.1), where the mesoscopic LBM model will fail. Another physically inconsistent result in Fig. 7(b) is that $|\lambda|/R$ based on the PLF force function decreases monotonically with decreasing gap for $n < 1.2$.

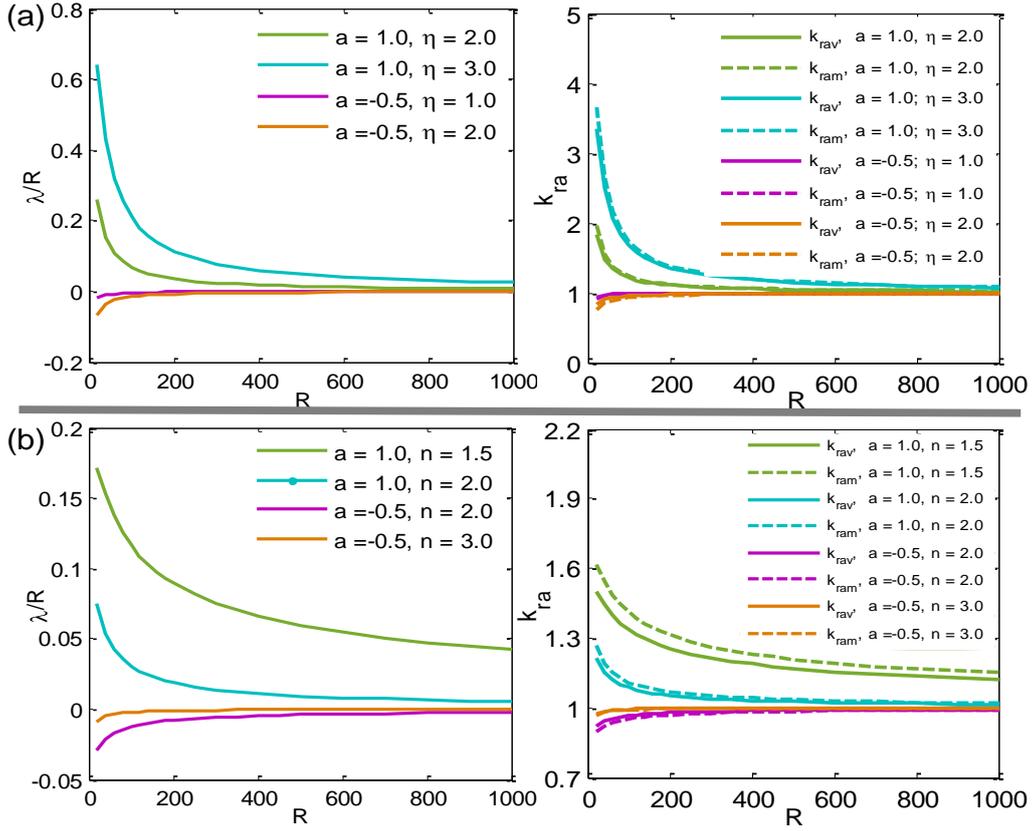

**Fig. 6.** Normalized slip length and permeability ratio as functions of the size $R$ of the gap in the parallel-plate channel based on (a) the EXF and (b) the PLF force functions, respectively.

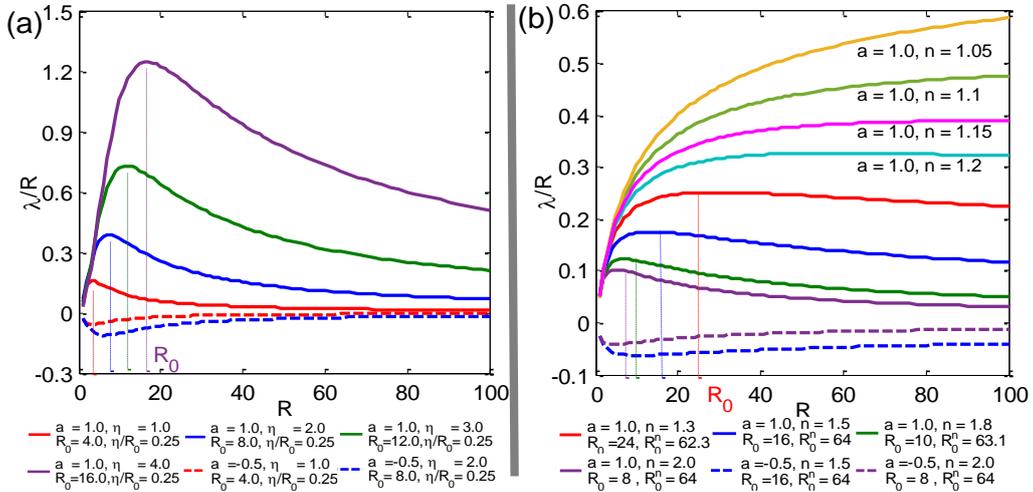

**Fig. 7.** Normalized slip length as a function of the size $R$ of the gap of the parallel-plate channel based on (a) the EXF and (b) the PLF force functions, respectively. Under each subpanel, the critical gap size $R_0$ for each set of interaction parameters for the two force functions is given based on the inequality (41). The relations $\eta/R_0 = 0.25$ and $R_0{}^n = 64$ are obtained for the EXF and PLF force functions, respectively.

To avoid these undesirable cases, the gap size for investigation should have $R > R_0$. The following critical gap sizes for the two force functions are obtained from Fig. 7:

$$R_0 = \eta/0.25, \tag{42a}$$
$$R_0 = 64^{1/n}. \tag{42b}$$

Moreover, $n \geq 1.2$ should be retained for the PLF force function. Therefore, to remain consistent with the increase in permeability ratio with decreasing pore size, the interaction parameters $\eta$ $(n)$ in the EXF (PLF) force function should satisfy

$$\eta < 0.25R, \tag{43a}$$
$$n > \max[1.2, \log_R 64]. \tag{43b}$$

Equations (39) and (43) provide useful guidelines for the reasonable range of interaction parameters, $(a, \eta)$ and $(a, n)$, when our derived mesoscale analytical solutions are used in experimental investigations.

### 4.3. Relationship between permeability ratios and dimensionless numbers

The flow-enhancement ratio of slip-flow systems is of great interest to the engineering community. Equations (37) and (38) provide exact but not straightforward expressions of this ratio as a function of the two dimensionless numbers. We thus propose exponential relationships between them for the EXF and PLF force functions, respectively,

$$k_{\mathrm{ra}} = \exp[\theta_0 (a\eta)^{\theta_1} (\eta/R)^{\theta_2}], \tag{44a}$$
$$k_{\mathrm{ra}} = \exp[\theta_0 (a/R^{n-1})^{\theta_1} (n-1)^{\theta_2}]. \tag{44b}$$

where $\theta_0$, $\theta_1$, and $\theta_2$ are unknown coefficients that can be determined by fitting. When any of the dimensionless numbers is zero, the permeability ratio approaches unity, which recovers the no-slip flow. Finally, note that the fluid slip ($k_{\mathrm{ra}} > 1$) is the focus of the investigation here.

In addition to the two curves shown in Fig. 4 and the two curves shown in Fig. 6, four additional curves within the reasonable range of interaction parameters are calculated from Eqs. (32) and (33). Based on these analytical results, the fitted surface using the relationship in Eq. (44) can be obtained with a good fitting quality, as shown in Fig. 8 for each of the force functions. Table 2 provides the fitted coefficients. In this sense, within the given range of dimensionless numbers, the relationships between permeability ratios and dimensionless numbers can be described in a straightforward way.

**Table 2** Coefficients for fitted permeability ratio from Eq. (44).

| Cases | | $\theta_0$ | $\theta_1$ | $\theta_2$ | $\theta_2/\theta_1$ |
|---|---|---|---|---|---|
| EXF force function | $k_{\mathrm{rav}}$ | 1.1197 | 1.4215 | 0.6840 | 0.4812 |
| | $k_{\mathrm{ram}}$ | 1.3644 | 1.2995 | 0.6846 | 0.5268 |
| PLF force function | $k_{\mathrm{rav}}$ | 852.33 | 1.8622 | 4.6409 | 2.4921 |
| | $k_{\mathrm{ram}}$ | 450.88 | 1.6957 | 4.1048 | 2.4207 |

It is interesting to observe from Table 2 that $\theta_2/\theta_1 \approx 0.5$ for the EXF force function, whereas $\theta_2/\theta_1 \approx 2.5$ for the PLF force function. By using these two approximations in Eq. (44), the pairs of dimensionless numbers, $(a\eta, \eta/R)$ and $(a/R^{n-1}, n-1)$, for each of the proposed force functions can be combined into a single dimensionless number, such as $a\eta^{1.5}/R^{0.5}$ and $a(n-1)^{2.5}/R^{n-1}$, respectively. In this way, the flow-enhancement ratio can be determined solely from the single dimensionless number, which serves as an indicator of the key feature of the slip flow system.

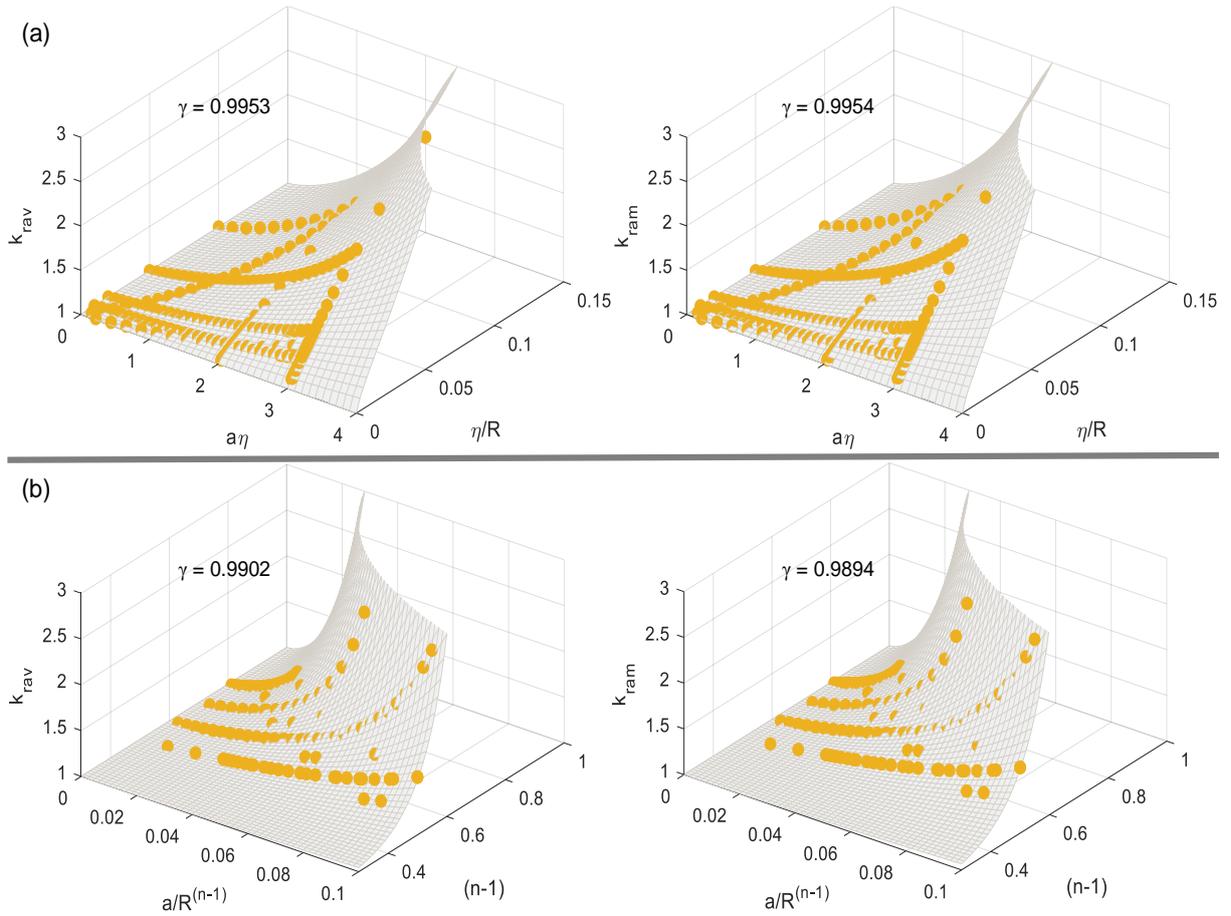

**Fig. 8.** Permeability ratio as a function of the two dimensionless numbers for (a) the EXF and (b) the PLF force functions, respectively. The circles are the analytical results from Eqs. (32) and (33), and the surfaces are based on the fitting functions [Eq. (44)]. In each case, the Pearson correlation coefficient $\gamma$ is given to show the quality of surface fitting.

### 4.4. Applications in a benchmark slip flow

A benchmark slip-flow experiment [5] was carried out in a rectangular duct [see Fig. 9(a)]. The length, width, and depth are aligned with the $x$, $y$, and $z$ coordinates, respectively, and the dimensions are $8.25\ \text{cm} \times 300\ \mu\text{m} \times 30\ \mu\text{m}$. The solid walls ($\alpha_1$, $\alpha_2$, $\beta_1$, and $\beta_2$) in the flow direction are uniformly hydrophilic or hydrophobic. The clean glass surface is naturally hydrophilic, whereas it becomes hydrophobic when covered with an octadecyl-trichlorosilane (OTS) layer. The deionized water is injected into the microchannel at a constant flux by using a syringe pump. The velocity profile, measured at $z = 15\ \mu\text{m}$ from the solid surface by using micron-particle-image velocimetry ($\mu$-PIV), was used to validate the mesoscopic LBM model based on the interaction force function at the particle level [26], which differs from the interaction force function between fluid particles and the flat wall proposed in this study.

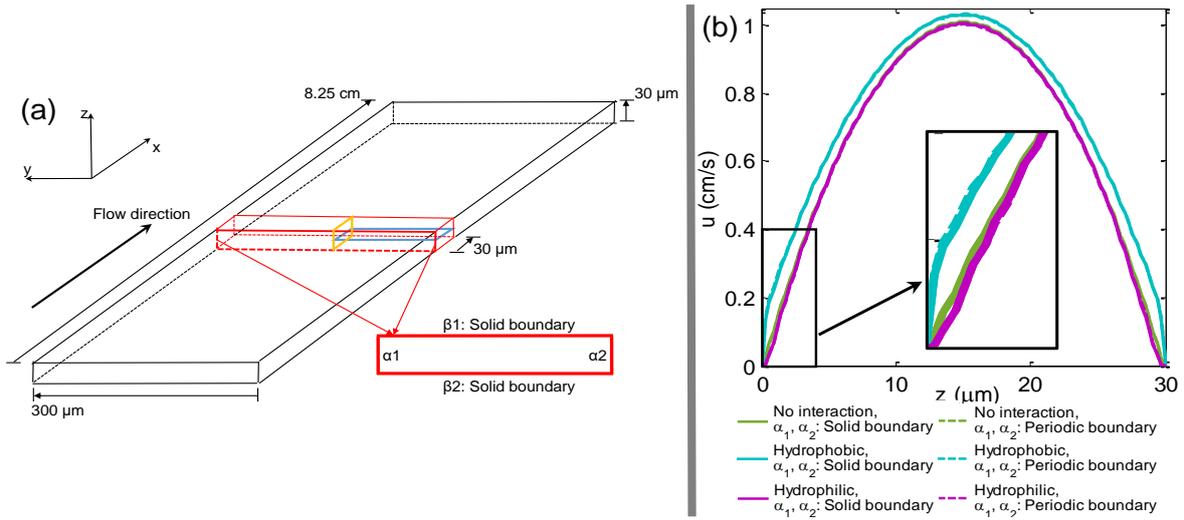

**Fig. 9.** (a) Dimensions of rectangular duct in slip-flow experiment of Ref. [5] (black lines) and in LBM simulation of Ref. [26] (red lines), respectively. The blue lines ($z = 15\ \mu m$) and yellow lines ($y = 150\ \mu m$) show the planes of investigation in these works and in this study, respectively. The red rectangle shows the cross section of the microchannel. The top and bottom walls, $\beta_1$ and $\beta_2$, are solid boundaries (where both the no-slip boundary condition and the F-S interaction force are applied), and the left and right walls, $\alpha_1$ and $\alpha_2$, are solid or periodic boundaries. (b) Comparison of the simulated velocity profiles using the validated LBM model when $\alpha_1$ and $\alpha_2$ are solid and periodic boundaries, respectively.

This section applies our analytical solutions for the parallel-plate microchannel to slip-flow experiments. Given that the aspect ratio of the cross section of the flow channel is $300\ \mu m$ : $30\ \mu m = 10 : 1$, the velocity profile of interest is at $y = 150\ \mu m$ in the vertical direction, which is controlled by the top and bottom walls ($\beta_1$ and $\beta_2$) with a narrow gap. This velocity profile was not measured in the flow experiment of Ref. [5] but can be obtained from the numerical simulations using the validated LBM model of Ref. [26]. Moreover, we should check whether the effect of the left and right walls ($\alpha_1$ and $\alpha_2$) is negligible, and then determine if the proposed analytical solutions (for two parallel walls) are applicable to the rectangular duct (with four solid walls) used in the flow experiment.

The walls $\beta_1$ and $\beta_2$ are always solid boundaries, whereas the walls $\alpha_1$ and $\alpha_2$ can be eliminated by being set up as periodic boundaries. Figure 9(b) shows that the velocity profiles for hydrophilic, hydrophobic, and no-interaction solid surfaces are almost the same regardless of whether $\alpha_1$ and $\alpha_2$ are solid boundaries or periodic boundaries. These results show that the rectangular duct (with four solid walls) in the flow experiment can be simplified as a parallel-plate microchannel in terms of the velocity profile over the vertical direction, where our analytical solutions (for two parallel walls) can be applied.

In our analytical method, $2R = 600$ lu, $\nu = 3.0\ lu^2/ts$, and $\rho_0 = 1.0\ mu/lu^3$ are used to represent the size of the gap in the flow channel ($30\ \mu m$), the kinematic viscosity of water ($1.5 \times 10^{-6}\ m^2/s$), and the density of water ($1.0 \times 10^3\ kg/m^3$), respectively. As illustrated in Fig. 10, by fitting the analytical velocity profiles based on the two proposed force functions to the simulated velocity profile over the vertical direction, the interaction parameters ($a, \eta$) and ($a, n$) can be obtained.

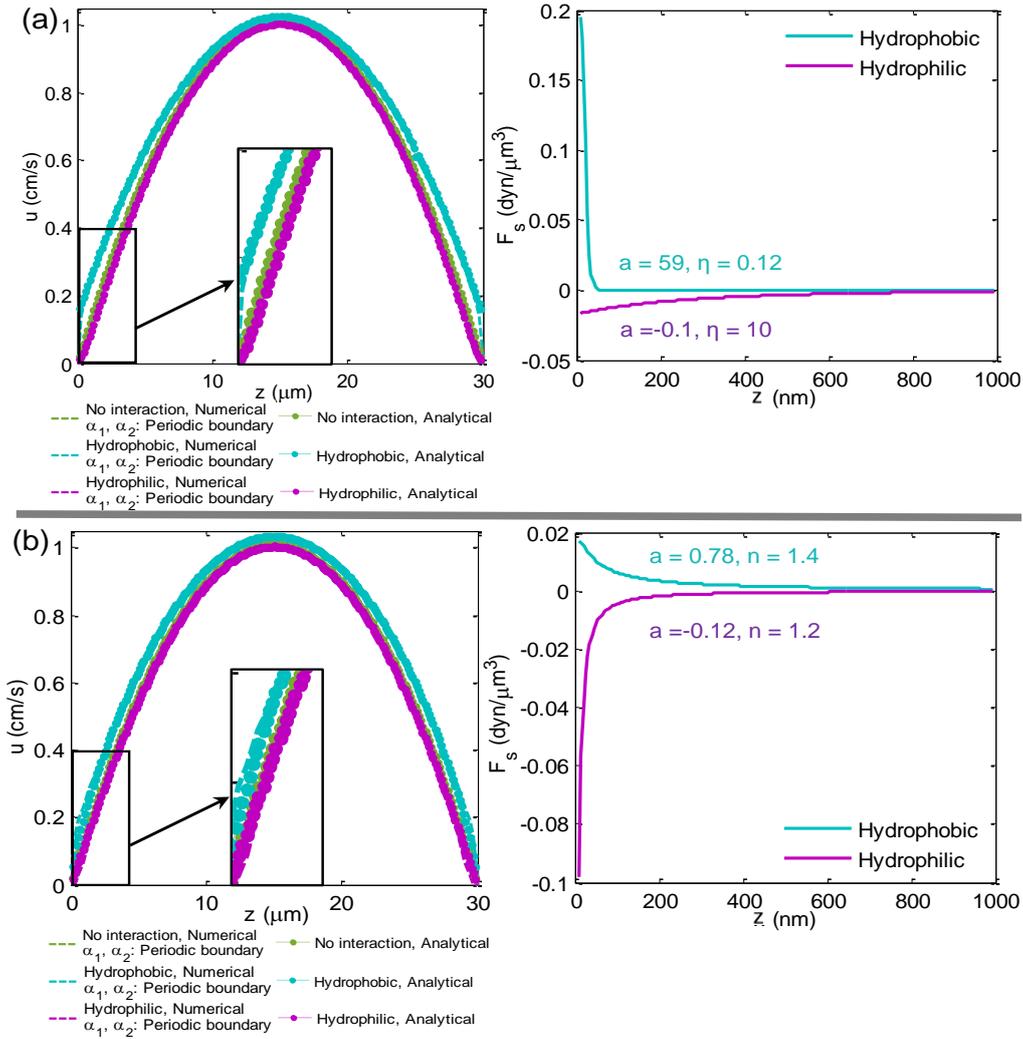

**Fig. 10.** Comparison of numerical and analytical velocity profiles (the numerical profile is obtained from the validated LBM model), and the recovered F-S interaction force based on (a) the EXF and (b) the PLF force functions, respectively.

The slip lengths calculated based on the EXF and PLF force functions are 1.16 and 1.22 μm, respectively, for a hydrophobic surface, which are consistent with the slip lengths around 1μm estimated by Tretheway and Meinhart [5]. For the hydrophilic surface, the calculated slip lengths are −0.37 and −0.30 μm, respectively, which were noticed but not quantified in the simulations of Zhu et al. [32]. The density ratios based on the EXF and PLF force functions are 0.998 and 0.182, respectively, for the hydrophobic surface, and they are 2.47 and 1.47 , respectively, for the hydrophilic surface. These density ratios are roughly within the physically reasonable range given by Eq. (39). The EXF and PLF force functions match the same velocity profiles on hydrophobic or hydrophilic surfaces but give different density ratios, implying different film fluid density behaviors of the two force functions. So far, the question remains open of whether the EXF or PLF function is more practical because accurate observations of the density profile near the solid surface and the density ratios for validating the interaction force functions remain unavailable.

Figure 10 shows the force curves for the two recovered force functions with the density profile included. To obtain the same velocity profile on a hydrophobic surface, the EXF force curve has a shorter interaction distance but a larger interaction strength, whereas the PLF force curve has a weaker interaction strength but a longer interaction distance. Similar findings can also be obtained on the hydrophilic surface. The recovered F-S force curve can be used to predict the slip flow for the given fluid and solid surface but with different gap sizes.

## 5. Conclusions

This paper proposes continuous force functions decaying (i) exponentially and (ii) by a power law to describe the physics of the F-S interaction force in the LBM model framework, which is responsible for liquid slip flow between two confined parallel plates and with a Knudsen number Kn = 0.001–0.1. Analytical solutions are derived by using both the macroscopic fluid pressure equation accounting for the mesoscale F-S interaction and the macroscopic NS equation recovered from the lattice Boltzmann equation. The analytical solutions for the density profile, velocity profile, slip length, and permeability ratio relate the macroscale flow properties to the mesoscale F-S interaction parameters and the size of the gap in the flow channel. Two dimensionless numbers (that is, $a\eta$ and $\eta/R$, or $a/R^{n-1}$ and $n-1$) are also derived for each slip-flow system based on the corresponding interaction force. They contribute to formulating the liquid flow in the slip regime in microfluidic devices and in tight porous media. These results should provide useful insights for petroleum engineers and soil scientists.

The analytical solutions with single integral or multiple integral included are numerically solvable by using MATLAB integration functions and are thoroughly validated by the LBM solutions. The F-S interaction parameters do not take arbitrary values. The physically reasonable ranges are determined by the observed range of density ratio (such as $0.1 < \rho_{\mathrm{ra}} < 2.0$), and the permeability ratio, which increases as the gap narrows (leading to $\eta < 0.25R$ and $n > \max[1.2, \log_R 64]$). The different behavior of film and bulk fluid density profiles and near-wall and bulk flow velocity profiles reflect the different decay forms of the two force functions. Within the given range of dimensionless numbers, simple relationships between permeability ratios and dimensionless numbers are obtained by fitting, and the two dimensionless numbers for each of the two force functions can be combined into a single dimensionless number for each force function, for example, $a\eta^{1.5}/R^{0.5}$ and $a(n-1)^{2.5}/R^{n-1}$ for the exponential and power-law decays, respectively.

We also apply the analytical solutions to a benchmark flow experiment that measured the slip velocity profile [5]. The analytical solutions are applicable because the rectangular duct can be simplified as a parallel-plate microchannel for dealing with the velocity profile in the vertical direction. The continuous F-S interaction force with two free interaction parameters were calibrated by fitting the velocity profile simulated by using the validated LBM model [26]. The proposed analytical model with the fitted F-S interaction force should nonetheless be further verified for a pair of solid surface and liquid but with different gap sizes. In future work, we plan to investigate which form of interaction force function is more suitable to the given solid surface using the available liquid density measurements.